\documentclass[notitlepage,aps,prl,twocolumn,superscriptaddress,amsmath,showpacs,tightenlines,longbibliography,10pt,footnote]{revtex4-1}

\usepackage[]{lineno}
\usepackage{amssymb}
\usepackage{amsmath}
\usepackage{dcolumn}
\usepackage{graphicx}
\usepackage{mathrsfs}
\usepackage{subfigure}
\usepackage{booktabs}
\usepackage{color}
\usepackage{docmute}
\usepackage{hyphenat}
\usepackage{CJKutf8}
\usepackage{multirow}
\usepackage{array}
\usepackage{threeparttable}
\usepackage[mathscr]{euscript}

\usepackage{graphicx}
\usepackage{subfigure}
\usepackage{tikz}
\usetikzlibrary{positioning}
\usepackage{diagbox}

\usepackage{ulem}
\newcommand{\bea}{\begin{eqnarray}}
\newcommand{\eea}{\end{eqnarray}}
\newcommand{\beq}{\begin{equation}}
\newcommand{\eeq}{\end{equation}}
\newcommand{\nn}{\nonumber}
\def\/{\over}

\usepackage{url}
\usepackage[colorlinks]{hyperref}
\hypersetup{%
	plainpages=true,
	breaklinks=true,
	hypertexnames=false,
	pageanchor=true,
	bookmarksnumbered=true,
	colorlinks=true,
	linkcolor={blue},
	citecolor={red},
	urlcolor={blue},
	anchorcolor={black}
}

\begin{document}
\title{Probing the Circular Unruh Effect  with Cavity-Controlled  Lamb Shifts}

\author{Yan Peng}
\affiliation{Department of Physics, Key Laboratory of Low Dimensional Quantum Structures and Quantum Control of Ministry of Education, and Hunan Research Center of the Basic Discipline for Quantum Effects and Quantum Technologies, Hunan Normal University, Changsha, Hunan 410081, China}
\affiliation{School of Fundamental Physics and Mathematical Sciences, Hangzhou Institute for Advanced Study, University of Chinese Academy of Sciences, No.1 Xiangshan Branch, Hangzhou 310024, China}

\author{Jiawei Hu}
\email[Corresponding author: ]{jwhu@hunnu.edu.cn}
\affiliation{Department of Physics, Key Laboratory of Low Dimensional Quantum Structures and Quantum Control of Ministry of Education, and Hunan Research Center of the Basic Discipline for Quantum Effects and Quantum Technologies, Hunan Normal University, Changsha, Hunan 410081, China}
\author{Hongwei Yu}
\email[Corresponding author: ]{hwyu@hunnu.edu.cn}
\affiliation{Department of Physics, Key Laboratory of Low Dimensional Quantum Structures and Quantum Control of Ministry of Education, and Hunan Research Center of the Basic Discipline for Quantum Effects and Quantum Technologies, Hunan Normal University, Changsha, Hunan 410081, China}

\begin{abstract}

The Unruh effect predicts that accelerated observers perceive the inertial vacuum as populated by particles, providing a flat-spacetime analogue of Hawking radiation. Its direct observation, however, remains experimentally challenging, since an Unruh temperature of $1\,\mathrm{K}$ requires accelerations of order $10^{20}\,\mathrm{m/s^2}$. Here, we show that the Lamb shift of a centripetally accelerated atom inside a high-$Q$ cavity provides a sensitive spectroscopic probe of the Unruh effect at dramatically lower accelerations. The cavity reshapes the electromagnetic density of states and converts otherwise tiny noninertial corrections into tunable level shifts.  Depending on the atomic angular velocity and cavity detuning, the Lamb shift can be  enhanced, strongly quenched, or completely screened.  Remarkably, for experimentally realistic parameters, a rotation-induced shift of order $10\;\mathrm{Hz}$ can arise already at accelerations as low as $0.5\,\mathrm{m/s^2}$, more than twenty orders of magnitude  below the acceleration scale conventionally associated with direct Unruh detection.  
These results identify cavity-controlled Lamb-shift spectroscopy as a viable route toward laboratory tests of the circular Unruh effect in the ultralow-acceleration regime.

\end{abstract}

\maketitle

\emph{Introduction}---The Unruh effect predicts that a uniformly accelerated observer perceives the vacuum defined by an inertial observer as a thermal bath with a temperature proportional to the proper acceleration~\cite{Davies75,Unruh76,Fulling73}. 
Through the equivalence principle, it is closely connected to Hawking radiation from black holes~\cite{Hawking74,Hawking75}, and is therefore widely regarded as a fundamental prediction at the interface of quantum theory, gravitation, and thermodynamics. 
Yet its direct observation remains elusive. A simple estimate shows that an acceleration of order $10^{20} \, \text{m/s}^2$ is required to produce an Unruh temperature of only $1 \text{K}$.  This enormous acceleration scale has long been regarded as the main obstacle to direct experimental tests of acceleration-induced vacuum effects. A central question is therefore whether signatures of the Unruh effect can be made observable at accelerations far below this conventional scale.

The Lamb shift~\cite{Lamb47} provides a natural probe of such effects. It originates from the coupling of atomic degrees of freedom to quantum vacuum fluctuations and has been measured with extraordinary precision. For example, hydrogen Lamb-shift spectroscopy has reached relative precision at the level of $10^{-6}$~\cite{Bezginov19}, making it sensitive to minute modifications of radiative level shifts. Since acceleration modifies the vacuum fluctuations sampled by an atom, the Lamb shift offers a direct spectroscopic window into noninertial quantum physics. Previous studies showed that uniform acceleration modifies the Lamb shift through the altered  vacuum fluctuations~\cite{Audretsch95A,Passante98}. In free space, however, these corrections are extremely small and remain far beyond current experimental reach, especially at low accelerations.

Centripetal acceleration offers an alternative route to noninertial quantum effects. An observer in circular motion through the inertial vacuum detects radiation, often referred to as the circular Unruh effect. Unlike the uniformly linearly accelerated case, however, the corresponding response is nonthermal~\cite{Letaw1980,Bell83,Hacyan1986,Bell87,Kim1987,Unruh98}. This qualitative distinction has motivated studies of circular-motion-induced modifications of atomic radiative processes and Lamb shifts~\cite{Audretsch95CQG,yu261}. Nevertheless, in free space, the resulting level shifts remain far too weak to be observed for experimentally accessible angular velocities. The challenge is thus not only to detect the circular Unruh effect, but to do so at accelerations many orders of magnitude below those usually associated with Unruh temperatures of order kelvin.

A natural strategy is to engineer the vacuum environment experienced by the accelerated atom. Optical and microwave cavities provide precisely this possibility.   
Through the Purcell effect~\cite{Purcell46}, cavity confinement reshapes the vacuum electromagnetic density of states, thereby modifying atom–field interactions.  
Related  works have shown that  cavities can amplify acceleration-induced modifications of vacuum fluctuations~\cite{Scully03,Lochan20,Stargen22,Arya23,Arya24,Zheng25,yu26,sahota2604}. 
However, cavity confinement does more than simply enhance radiative effects: by changing the spectral weight sampled by the atom, it can also suppress or cancel vacuum-induced level shifts. This raises the central question addressed here: can cavity control make acceleration-induced Lamb shifts observable at ultralow accelerations?

Here we show that the answer is affirmative. We investigate the Lamb shift of a centripetally accelerated atom in a high-quality cavity, and demonstrate that cavity confinement produces strong,  tunable, and spectroscopically resolvable signatures of acceleration-modified vacuum fluctuations.  Depending on the cavity detuning and angular velocity, the Lamb shift can be enhanced, strongly quenched, or completely screened.  
 In the ultralow-acceleration regime, the noninertial correction scales as $Q^3\Omega^2/\omega_0^2$, allowing a rotation-induced shift of order $10\;\mathrm{Hz}$, which is within the resolution of current spectroscopic techniques,  already at an acceleration of $0.5\;\mathrm{m/s^2}$. This is more than twenty orders of magnitude below the acceleration scale associated with a $1\;\mathrm{K}$ Unruh temperature. 
At higher accelerations (of order $\sim 10^{6}\,\mathrm{m/s^2}$), rotation-assisted sideband resonances make the Lamb shift dominated by noninertial vacuum fluctuations. Our results show that cavity-controlled Lamb-shift spectroscopy can bring the circular Unruh effect into the ultralow-acceleration regime accessible to precision measurements.

\emph{General expression for the Lamb shift}---We consider a two-level atom undergoing uniform centripetal acceleration inside a cavity and interacting with electromagnetic vacuum fluctuations. The atom possesses a proper transition frequency $\omega_0$ and follows the trajectory $x(t)=R\cos(\Omega t),~y(t)=R\sin(\Omega t), ~z(t)=0$, where $R$ is the orbital radius, $\Omega$  the angular velocity, and $t$  the laboratory (coordinate) time. In the laboratory frame, the atom–field interaction Hamiltonian can be organized in the following form
: $H_{I}=-\sum_{m=\rho,\phi,z}D_{m} \mathcal{E}_{m}$, where $D_{m}$ denotes the atomic electric dipole moment operator, and $\mathcal{E}_m$ are defined as
\begin{eqnarray}
\mathcal{E_{\rho}}&=&\Omega R B_{z} +\cos(\Omega t) E_{x} +\sin(\Omega t) E_y,\nonumber\\
\mathcal{E_{\phi}}&=&[\cos(\Omega t)  E_{y} - \sin(\Omega t) E_{x}]/\gamma, \nonumber\\
\mathcal{E}_{z}&=&E_{z}   -\Omega R \sin(\Omega t)B_{y}  -\Omega R \cos(\Omega t)B_{x},
\end{eqnarray}
with $\gamma=\left(1-{\Omega^2 R^2}/{c^2}\right)^{-\frac{1}{2}}$ being the Lorentz factor,  $c$ the speed of light, and  $E_i$, $B_i$ $(i=x,y,z)$ the electric and magnetic field components in the laboratory frame.

In the framework of open quantum systems~\cite{breuer07}, the Lamb-shift correction to the atomic energy level spacing in the laboratory frame can be expressed as
\begin{align}\label{shift}
  \Delta =&\sum_{\alpha ,\beta =\rho ,\phi ,z}{\frac{d_{\alpha}d_{\beta}^{*}}{2\pi \hbar^2}}\nn\\
  & \times \text{P}\int_{-\infty}^{\infty}{\text{d}}\nu \;\mathcal{G}_{\alpha \beta}\left( \nu \right) \left( \frac{1}{\nu +\omega }-\frac{1}{\nu -\omega } \right) \;,
\end{align}
where $d_\alpha = \langle e | D_\alpha | g \rangle$ is the transition matrix element of the dipole operator $D_\alpha$ between the excited state $|e\rangle$ and ground state $|g\rangle$, $\hbar$ is the reduced Planck constant, and $\text{P}$ denotes the Cauchy principal value. The quantity $\omega=\omega_0/\gamma$ is the transition frequency in the laboratory frame, and $\mathcal{G}_{\alpha \beta} (\nu)$ denotes the Fourier transform of the field correlation function,
\begin{equation}\label{Gtwofuction}
   \mathcal{G}_{\alpha \beta} (\nu)= \int_{-\infty}^{\infty} \text{d}t_{-} e^{i \nu t_-}G_{\alpha \beta}\left( t_- \right) \;,
\end{equation}
where  $t_{-} = t - t'$ and $G_{\alpha\beta}(t_{-}) = \langle 0 | \mathcal{E}_{\alpha}(t, \mathbf{x}) \mathcal{E}_{\beta}(t', \mathbf{x}') | 0 \rangle$ represents the two-point correlation function of the electromagnetic field in the vacuum state $|0\rangle$. 
Owing to stationarity,  it depends only on the time difference $t_{-}$.
These two-point functions are determined by the cavity density of states $\rho(\omega_k)$, where $\omega_k$ denotes the field mode frequency. For a high-quality cavity, the density of states is typically described by a Lorentzian distribution centered at the cavity’s normal mode frequency $\omega_c$, i.e., $
    \rho(\omega_k) = \frac{1}{\pi} \frac{\omega_c / Q}{(\omega_c / Q)^2 + (\omega_k - \omega_c)^2}$,
with $Q$ representing the cavity quality factor. Throughout this Letter, we assume $Q \gg 1$. 
Explicit expressions of the corresponding two-point functions $\langle 0 | \mathcal{E}_\alpha(t, \mathbf{x}) \mathcal{E}_\alpha(t', \mathbf{x}') | 0 \rangle$ are presented in the End Matter.

In the nonrelativistic regime, where the linear velocity $v = R\Omega$ is much smaller than the speed of light $c$ ($v\ll c$), keeping only the leading-order terms, the Lamb shift in the laboratory frame becomes
\begin{align}\label{Lambshift}
    &\Delta= \frac{d_z^2}{6 V \epsilon _0 \hbar }\left[\mathcal{F}\left(-\omega_0 \right)-\mathcal{F}\left(\omega_0 \right)\right]+\frac{d_{\rho }^2+d_{\phi }^2}{12 V \epsilon _0 \hbar }\left[\mathcal{F}\left(-\omega_0 -\Omega \right)\right.\nonumber\\
    &\left.-\mathcal{F}\left(\omega_0 +\Omega \right)-\mathcal{F}\left(\omega_0 -\Omega \right)+\mathcal{F}\left(\Omega -\omega_0 \right)\right]+\mathcal{O}[({R\Omega}/{c})^2]\;,
\end{align}
where
\begin{align}\label{Fomegak}
    \mathcal{F}(\omega_k)=&\left\{\left(\arctan Q+\frac{\pi }{2}\right)\left[Q (\omega_{c}-\omega_k)+\frac{\omega_{c}}{Q}\right]\right.\nonumber\\
    &\left.+\frac{\omega_k}{2}  \ln \left[\left(1+\frac{1}{Q^2}\right)\frac{ \omega_{c}^2}{\omega_k^2}\right]\right\}\rho (\omega_k) \;.
\end{align}
Here, $\epsilon_0$ is the vacuum permittivity and $V$ is the cavity volume; higher-order terms $\mathcal{O}[(R\Omega/c)^{2}]$ are neglected. 
Equation~(\ref{Lambshift}) displays the key physical mechanism. 
To leading order in $R\Omega/c$, the axial contribution depends only on $\mathcal{F}(\pm\omega_0)$ and is therefore unchanged by circular motion. By contrast, the transverse sector contains the shifted frequencies $\omega_0\pm\Omega$ and $\Omega-\omega_0$. Thus, the rotation-induced Lamb shift is governed by transverse vacuum fluctuations and their coupling to cavity sidebands.

As a baseline, we consider the inertial limit $\Omega=0$, for which the Lamb shift reduces to
\begin{align}\label{Lambshiftin}
\Delta_{\rm inertial}=\frac{d_z^2+d_{\rho }^2+d_{\phi }^2}{6 V \epsilon _0 \hbar }\left[\mathcal{F}\left(-\omega_0 \right)-\mathcal{F}\left(\omega_0 \right)\right]\;.
\end{align}
In this case, the contributions from different polarization components are identical. This provides a reference for analyzing the effects of rotation.  
The noninertial correction due to circular motion is 
\begin{align}\label{Lambshiftcir}
   &\Delta_{\rm circular} =  \Delta-\Delta_{\rm inertial}\nn\\
   &\;\;\;\;=\frac{d_{\rho }^2+d_{\phi }^2}{12 V \epsilon _0 \hbar }\left[\mathcal{F}\left(-\omega_0 -\Omega \right)-\mathcal{F}\left(\omega_0 +\Omega \right)-\mathcal{F}\left(\omega_0 -\Omega \right)\right.\nonumber\\
   &\;\;\;\;\;\;\;\;\;\left.+\mathcal{F}\left(\Omega -\omega_0 \right)-\mathcal{F}\left(-\omega_0 \right)+\mathcal{F}\left(\omega_0 \right)\right]+\mathcal{O}[({R\Omega}/{c})^2]\;.
\end{align}
Thus, circular motion modifies the Lamb shift through transverse polarization alone at leading order. This is why isotropically polarizable atoms exhibit behavior that is absent in axial-only models.

\emph{Cavity-controlled Lamb shifts of centripetally accelerated atoms}---Building on the expressions derived above, 
we examine how the Lamb shift of a centripetally accelerated atom evolves with increasing rotation and identify the angular velocity regimes in which rotation-induced changes become experimentally observable.

\emph{a. Inertial baseline}---When the cavity’s normal mode frequency $\omega_c$ is tuned near the atomic transition frequency, the Lamb shift of an inertial atom can be significantly enhanced, reaching extrema at $\omega_c\approx\omega_0 (1\pm1/Q)$. The corresponding magnitude is well approximated by
\begin{align}\label{inertial-omega-big}
    \Delta_{\rm inertial}\approx \mp\frac{ Q }{12 V \epsilon _0 \hbar }\left(d_{\rho }^2+d_{\phi}^2+d_{z }^2\right)\;.
\end{align}
Interestingly, the Lamb shift can also be completely quenched. By continuity, the sign change of $\Delta$ implies the existence of a zero crossing at an intermediate cavity frequency between $\omega_{c} \approx\omega_{0}(1-1/Q)$ and $\omega_{c} \approx\omega_{0}(1+1/Q)$. 
At this point, the vacuum-fluctuation-induced corrections to the ground and excited states are completely canceled, resulting in a vanishing relative Lamb shift.

\emph{b. Significant rotational suppression at ultralow angular velocity}
---We now consider the regime of small angular velocity, $0<\Omega\ll\omega_0/Q$.  A straightforward asymptotic analysis shows that the rotation-induced correction reaches an extremum when the cavity is tuned to $\omega_c\approx\omega_0[1 \pm (\sqrt{2}-1)/{Q}]$, for which
\begin{align}\label{circular-small-suppression}
    \Delta_{\rm circular} \approx \pm \frac{ (3+2 \sqrt{2}) Q^3 \Omega ^2 }{48 V \epsilon _0 \hbar \omega _0^2 } (d_{\rho }^2+d_{\phi }^2)\;.
\end{align}
For reference, the inertial Lamb shift at the same cavity detuning is 
\begin{align}\label{inertial-small}
    \Delta_{\rm inertial} \approx \mp \frac{  Q \left(d_z^2+d_{\rho }^2+d_{\phi }^2\right)}{12 \sqrt{2} V \epsilon _0 \hbar }\;,
\end{align}
which has the opposite sign. Consequently, at low angular velocities, rotation suppresses the Lamb shift. Although $\Omega$ is small in Eq.~\eqref{circular-small-suppression}, the rotational correction is parametrically enhanced by the quality factor $Q$, yielding $\Delta_{\rm circular}\propto Q^3\Omega^2/\omega_0^2$. This indicates that rotation can substantially reduce the total Lamb shift and allow resolvable noninertial shifts even at ultralow angular velocities.
For example, for an isotropically polarizable atom ($d_\rho^2 = d_\phi^2 = d_z^2 \equiv d^2$) with parameters consistent with the Markov approximation~\cite{footnote}: $d=10^{-29}\ \mathrm{C m}, \; V=10^{-10}\ \mathrm{m}^3, \;Q=10^7, \;R=50\ \mathrm{nm},$$ \;\omega_0=5\times10^{11}\ \mathrm{Hz} $, and an angular velocity $\Omega=3.1\times10^3\,\mathrm{rad/s}$ (corresponding to an approximate centripetal acceleration of $0.5 \,\mathrm{m/s^2}$), the inertial contribution still dominates the total level shift, but the rotation-induced correction reaches $\sim10\,\mathrm{Hz}$, well within current spectroscopic resolution.  Thus, noninertial signatures can be resolved without large accelerations through cavity amplification of the rotation-induced suppression of the Lamb shift.

\emph{c. Significant rotational enhancement at ultralow angular velocity}
---In the low-angular-velocity regime $0<\Omega\ll\omega_0/Q$, the rotation-induced Lamb shift also exhibits extrema when the cavity frequency is tuned to $\omega_c\approx\omega_0[1 \pm (\sqrt{2}+1)/{Q}]$, with a magnitude of
\begin{align}\label{circular-small-enhancement}
    \Delta_{\rm circular} \approx \mp \frac{ \left(3-2 \sqrt{2}\right) Q^3 \Omega ^2 }{48 V  \epsilon _0 \hbar \omega _0^2}\left(d_{\rho }^2+d_{\phi }^2\right)\;.
\end{align}
At the same detuning, the inertial Lamb shift remains as in Eq.~\eqref{inertial-small}. Here, the rotation-induced contribution has the same sign as the inertial Lamb shift and enhances the total level shift. As in the suppression regime discussed above, the rotational correction retains the characteristic $Q^3\Omega^2/\omega_0^2$ scaling. Consequently, the same cavity-enhanced mechanism that produces rotational suppression can also generate experimentally resolvable enhancements at ultralow angular velocities. 
For illustration, setting the angular velocity to $\Omega = 2\times 10^4 \,\mathrm{rad/s}$ (corresponding to a centripetal acceleration of $10 \,\mathrm{m/s^2}$) while keeping all other parameters identical to those chosen in Sec.~b, the rotation-induced contribution increases the Lamb shift by as much as $10 \;\mathrm{Hz}$. Although the total shift is still dominated by inertial effects, this rotational enhancement is well within current high-precision spectroscopic sensitivities. 
These results demonstrate that, analogous to rotation-induced suppression, signatures of rotation-induced enhancement are also resolvable at ultralow accelerations.

\emph{d. Rotation-dominated Lamb shifts beyond the ultralow-angular-velocity regime}---As the angular velocity increases to the regime $\omega_0/Q \ll \Omega \ll Q\omega_0$, tuning the cavity's normal mode frequency $\omega_c$ near the rotation-induced sidebands $|\omega_0 \pm \Omega|$ gives rise to a qualitatively new regime in which the Lamb shift is dominated by rotational effects. Specifically, pronounced resonances always occur at $\omega_c\approx(\omega_0+\Omega)(1\pm1/Q)$. Away from the commensurate point $\Omega=2\omega_0$, additional resonances emerge at $\omega_c\approx(\omega_0-\Omega)(1\pm1/Q)$ for $\Omega<\omega_0$, and at $\omega_c\approx(\Omega-\omega_0)(1\pm1/Q)$ for $\Omega>\omega_0$. 
Near these resonances, the Lamb shift is approximately
\begin{align}\label{Strong-enhancement}
\Delta \approx \frac{\eta Q}{24 V \epsilon_0 \hbar} (d_\rho^2 + d_\phi^2)\;,
\end{align}
where $\eta = +1$ for $\omega_c\approx(\omega_0\pm\Omega)(1-1/Q)$ or $\omega_c\approx(\Omega-\omega_0)(1+1/Q)$, corresponding to an increase in the atomic level spacing, and $\eta = -1$ for $\omega_c\approx(\omega_0\pm\Omega)(1+1/Q)$ or $\omega_c\approx(\Omega-\omega_0)(1-1/Q)$, corresponding to a decrease in the atomic level spacing. 
In this regime, the Lamb shift of the centripetally accelerated atom reaches the same order of magnitude as the maximum inertial Lamb shift in the same cavity, as given in Eq.~\eqref{inertial-omega-big}. 
Remarkably, the shift is determined entirely by the transverse polarization sector and scales linearly with the cavity quality factor $Q$. Depending on the resonance condition, the rotational contribution can be either positive or negative, allowing the total Lamb shift to be tuned between large positive and negative values. 
For comparison, the Lamb shift of an inertial atom in the same cavity is
\begin{align}\label{Strong-enhancement-inertial}
\Delta_{\rm inertial} \approx \frac{\omega_c \omega_0}{3 V \epsilon_0 \hbar (\omega_0^2 - \omega_c^2)} (d_\rho^2 + d_\phi^2 + d_z^2)\;.
\end{align}
By comparing the absolute-value ratio of Eqs.~\eqref{Strong-enhancement} and \eqref{Strong-enhancement-inertial}, one finds that, for an isotropically polarizable atom, the Lamb shift of a rotating atom exceeds that of its inertial counterpart by a factor of $\frac{Q (\omega_0^2 - \omega_c^2)}{12 \omega_c \omega_0}$. This result indicates that rotation drives the Lamb shift substantially away from its inertial value. Thus, in the sideband-resonant regime, rotation is no longer a perturbative correction but instead becomes the dominant source of the observed level shift. 
As an illustrative example, choosing $\Omega=5\times10^6~\mathrm{rad/s}$, corresponding to a centripetal acceleration of $\sim10^6~\mathrm{m/s^2}$, and tuning the system to the resonance $\omega_c\approx(\omega_0-\Omega)(1-1/Q)$ 
yields a Lamb shift of approximately $920~\mathrm{Hz}$, whereas the corresponding inertial Lamb shift is only $53~\mathrm{Hz}$. The observed level shift is therefore overwhelmingly dominated by rotational effects.

\emph{e. Rotation-induced quenching near the atomic resonance}---As the angular velocity increases to the regime $\Omega \gg \omega_0/Q$ and sufficiently detuned from the commensurate point $\Omega=2\omega_0$, tuning the cavity mode frequency $\omega_c$ near the atomic resonance $\omega_0$ leads to a qualitatively different behavior from the sideband-enhanced regime discussed above. In this regime, the Lamb shift of a centripetally accelerated atom is substantially suppressed relative to its inertial counterpart. Specifically, the extrema appear at $\omega_c \approx \omega_0(1 \pm 1/Q)$, with peak values
\begin{align}
\Delta \approx \mp \frac{ Q}{12 V \epsilon_0 \hbar} d_z^2\;.
\end{align}
For the same cavity tuning, an inertial atom reaches extrema at the same resonance,
\begin{align}
\Delta_{\rm inertial} \approx \mp \frac{ Q}{12 V \epsilon_0 \hbar} (d_\rho^2 + d_\phi^2 + d_z^2)\;.
\end{align}
Thus, rotation preserves both the resonance positions and sign pattern but redistributes the contributions from different polarization components. For an isotropically polarizable atom, the rotational peak is reduced by a factor of three relative to the inertial value, demonstrating a clear quenching of the cavity-enhanced Lamb shift near $\omega_c = \omega_0$.

An even more striking effect arises under the commensurate condition $\Omega=2\omega_0$. Near $\omega_c\approx\omega_0(1+1/Q)$, the Lamb shift becomes
\begin{align}\label{Delta_commensurate}
\Delta\approx\frac{Q }{24 V \epsilon_0 \hbar }\left(d_{\rho }^2+d_{\phi }^2-2 d_z^2\right)+\frac{d_{\rho }^2+d_{\phi }^2}{16 V \epsilon _0 \hbar }\;.
\end{align}
For isotropic polarization, this expression reduces to $\Delta\approx\frac{d^2}{8 V \epsilon _0 \hbar }$, reflecting a near cancellation between the transverse and axial polarization contributions. Compared with the inertial counterpart Eq.~\eqref{inertial-omega-big} under the same conditions, the Lamb shift is suppressed by a factor of order $Q$. A similar suppression occurs near $\omega_c\approx\omega_0(1-1/Q)$. These results indicate that the cavity-enhanced resonance peaks characteristic of inertial atoms near $\omega_c\approx\omega_0(1\pm1/Q)$ can be substantially quenched by rotational motion. At the commensurate point $\Omega=2\omega_0$, the quenching becomes particularly pronounced due to the near cancellation between transverse and axial polarization contributions, demonstrating that rotation can dramatically suppress cavity-enhanced vacuum-fluctuation effects and almost completely remove the cavity-enhanced Lamb shift near resonance.

\begin{figure}[!t]
    \centering
    \subfigure[ \;Rotation-induced Lamb shift $\Delta_{\rm circular}$]
    {\includegraphics[scale=0.37]{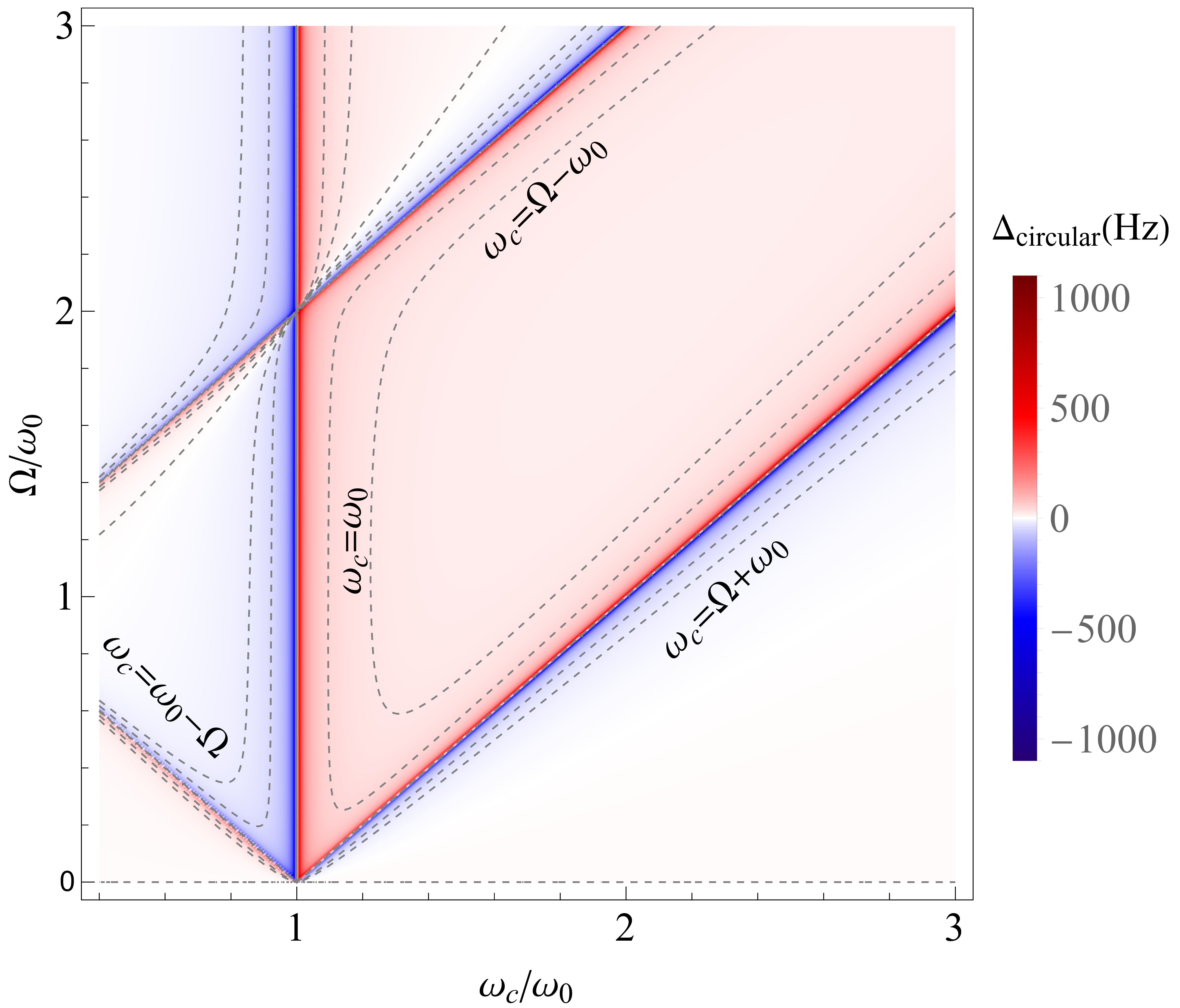}}
    \subfigure[ \;Total Lamb shift $\Delta$]{\includegraphics[scale=0.37]{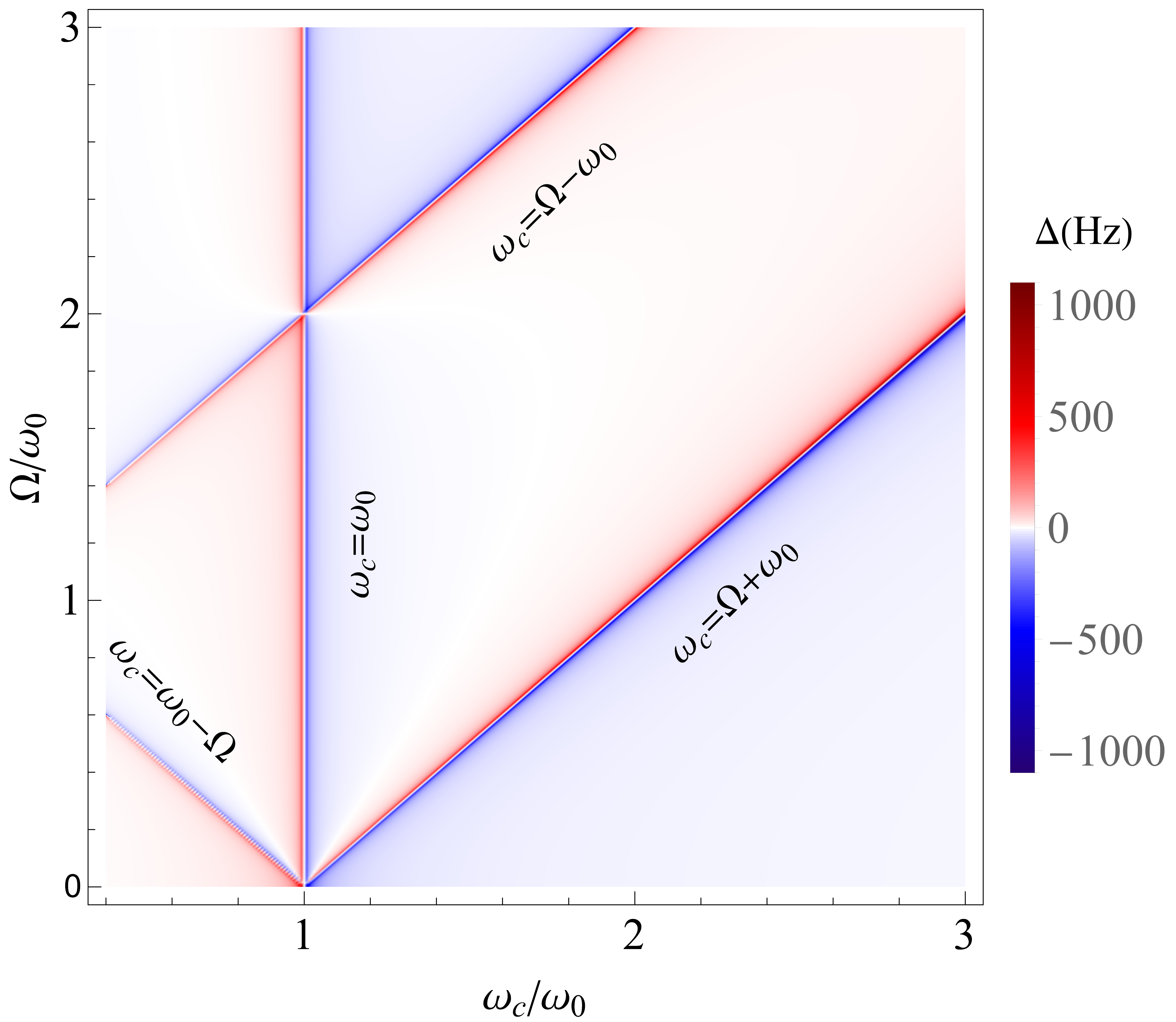}}
\caption{Density plots of (a) the rotation-induced Lamb shift $\Delta_{\rm circular}$ and (b) the total Lamb shift $\Delta$ as functions of $\omega_c/\omega_0$ and $\Omega/\omega_0$. Calculations assume an isotropically polarizable atom  with $d=10^{-29}\ \mathrm{C m}, \; V=10^{-10}\ \mathrm{m}^3$ and $Q=10^7$.}
  \label{Plotdensity}
\end{figure}

\emph{f. Rotation-induced screening of the Lamb shift}---An even more intriguing phenomenon is the complete screening of the Lamb shift induced by the interplay between rotation and cavity confinement, which is distinct from the peak suppression discussed above. While screening of the cavity Lamb shift can also occur for inertial atoms, it is typically restricted to the vicinity of the atomic resonance, $\omega_c\approx\omega_0$. For a rotating atom, by contrast, screening can arise not only near $\omega_c\approx\omega_0$ but also in the vicinity of the rotation-induced resonances $\omega_c\approx|\omega_0\pm\Omega|(1\pm1/Q)$. Since the Lamb shift exhibits opposite signs on the two neighboring resonant branches, continuity implies the existence of an intermediate angular velocity at which it vanishes exactly, yielding $\Delta=0$. For example, a complete cancellation occurs between the neighboring resonances $\omega_c\approx(\omega_0+\Omega)(1-1/Q)$ and $\omega_c\approx(\omega_0+\Omega)(1+1/Q)$. 
At this point, the vacuum-fluctuation contribution to the atomic level spacing vanishes entirely. This zero is a genuine consequence of the combined effects of rotation and cavity confinement, since no corresponding zero appears for an inertial atom in the same cavity over that interval. The resulting screening provides a striking manifestation of how rotational motion can qualitatively reshape vacuum-induced radiative corrections.

\emph{Discussion}---To provide an intuitive picture of how rotational motion modifies the Lamb shift, Fig.~\ref{Plotdensity} (a) displays the rotation-induced contribution in the $(\omega_c/\omega_0,\Omega/\omega_0)$ plane. The color scale indicates the magnitude of the deviation from the inertial value. At low angular velocities, the Lamb shift remains close to the inertial result,  except near the sharply enhanced cavity resonance around $\omega_c\approx\omega_0$. As $\Omega$ increases, a qualitatively new structure emerges: additional resonant branches appear approximately along 
$\omega_c\approx\omega_0-\Omega$ and $\omega_c\approx\Omega\pm\omega_0$, 
corresponding to the rotation-induced sideband resonances discussed above. The resonance positions shift with increasing $\Omega$, indicating that strong rotational effects require simultaneous tuning of both the cavity resonance frequency and the atomic angular velocity. Notably, a strong suppression occurs near $(\omega_c/\omega_0,\Omega/\omega_0)=(1,2)$, corresponding to the commensurate condition $\Omega=2\omega_0$, in agreement with the analytical prediction. Figure~\ref{Plotdensity}(b) further reveals zero-crossing lines of the total Lamb shift near both the ordinary atomic resonance $\omega_c \approx \omega_0$ and the rotation-induced sideband resonances $\omega_c\approx|\omega_0\pm\Omega|$, along which the Lamb shift is completely screened. The emergence of such screening structures demonstrates that the combined effects of rotation and cavity confinement can not only enhance or suppress vacuum-fluctuation corrections, but also eliminate them entirely.

Our results also clarify the role of atomic polarization. Previous studies of cavity-modified Lamb shifts under centripetal acceleration~\cite{Arya23} focused on atoms polarizable only along the rotation axis. Here, we treat the atom as isotropically polarizable, which is more relevant for realistic spectroscopic systems. We find that the transverse polarization sector plays a crucial role in the rotation-induced Lamb shift and gives rise to several qualitatively and quantitatively new phenomena absent in axial-only treatments. 
In the ultralow-angular-velocity regime, transverse polarization enables experimentally detectable rotational corrections even when the total Lamb shift remains close to its inertial value. For example, at an acceleration of only $0.5\;\mathrm{m/s^2}$, the acceleration-induced transverse contribution reaches the $\sim 10\;\mathrm{Hz}$ level, whereas the corresponding axial contribution is only $\sim 10^{-8}\;\mathrm{Hz}$. Thus, in the ultralow-acceleration regime, the transverse sector dominates the axial contribution by a factor of order $c^2/(\omega_0^2R^2)$, which is about $10^{9}$ for the parameters considered here. 
Near $\omega_c\approx\omega_0$, by contrast, the transverse contributions are shifted away from resonance while the axial contribution remains nearly unchanged, leading to substantial quenching of the Lamb shift. An even stronger quench occurs at the commensurate point $\Omega=2\omega_0$, where transverse and axial contributions are comparable in magnitude but opposite in sign, driving the total shift toward near cancellation.

\emph{Summary}---We have shown that the Lamb shift of a centripetally accelerated atom in a high-$Q$ cavity provides a sensitive probe of acceleration-modified vacuum fluctuations. Cavity confinement converts weak noninertial corrections into tunable spectroscopic signals: depending on the cavity detuning and angular velocity, the Lamb shift can be enhanced, strongly quenched, or completely screened.   In the ultralow-acceleration regime, the rotation-induced contribution scales as $Q^3\Omega^2/\omega_0^2$, allowing shifts of order $10\,\mathrm{Hz}$,  within current spectroscopic resolution,  already at accelerations of order $0.5\,\mathrm{m/s^2}$. This is more than twenty orders of magnitude below the acceleration scale associated with a $1\,\mathrm{K}$ Unruh temperature.   At higher accelerations, of order $\sim10^{6}\,\mathrm{m/s^2}$, rotation-assisted sideband resonances make the Lamb shift dominated by noninertial vacuum fluctuations, while near the atomic resonance rotation can suppress or cancel the cavity-enhanced inertial shift.  Remarkably, the peak noninertial shift can reach a magnitude comparable to the cavity-enhanced inertial Lamb shift in the same setup.  These results demonstrate that cavity-controlled Lamb-shift spectroscopy can dramatically lower the acceleration threshold for observing noninertial vacuum effects, opening a promising route toward laboratory tests of the circular Unruh effect.

\begin{acknowledgments}
\emph{Acknowledgments}---This work was supported in part by the NSFC under Grants No. 12075084 and 12575051, and the innovative research group of Hunan Province under Grant No. 2024JJ1006.
\end{acknowledgments}

\newpage

\section*{End Matter}

\setcounter{equation}{0}
\renewcommand{\theequation}{A\arabic{equation}}

\label{Appendix A}\emph{Electromagnetic-field correlation functions inside a cavity}---The Lamb shift is determined by the vacuum two-point functions of the electric field, $G_{\alpha\beta}(t-t^\prime)=\langle 0|\mathcal{E}_{\alpha}\left(t,\mathbf{x}\right)\mathcal{E}_{\beta}\left(t',\mathbf{x}'\right)|0\rangle$. For real dipole matrix elements $d_\alpha$, the cross terms in Eq.~(2) vanish, so that only the diagonal components contribute. The corresponding cavity correlation functions read
\begin{widetext}
\begin{eqnarray}
\langle 0|\mathcal{E}_{\rho}\left(t,\mathbf{x}\right)\mathcal{E}_{\rho}\left(t',\mathbf{x}'\right) |0\rangle
&=& \Omega^2 R^2\langle 0|B_{z}\left(t,\mathbf{x}\right)B_{z}\left(t',\mathbf{x}'\right) |0\rangle+ \Omega R \cos(\Omega t')\langle 0|B_{z}\left(t,\mathbf{x}\right)E_{x}\left(t',\mathbf{x}'\right)|0\rangle\nonumber\\
&&+ \Omega R \sin(\Omega t')\langle 0|B_{z}\left(t,\mathbf{x}\right)E_{y}\left(t',\mathbf{x}'\right) |0\rangle+ \Omega R \cos(\Omega t)\langle 0|E_{x}\left(t,\mathbf{x}\right)B_{z}\left(t',\mathbf{x}'\right) |0\rangle\nonumber\\
&&+ \cos(\Omega t)\cos(\Omega t')\langle 0|E_{x}\left(t,\mathbf{x}\right)E_{x}\left(t',\mathbf{x}'\right)|0\rangle
+ \cos(\Omega t)\sin(\Omega t')\langle 0|E_{x}\left(t,\mathbf{x}\right)E_{y}\left(t',\mathbf{x}'\right) |0\rangle\nonumber\\
&&+ \Omega R \sin(\Omega t)\langle 0|E_{y}\left(t,\mathbf{x}\right)B_{z}\left(t',\mathbf{x}'\right) |0\rangle
+ \sin(\Omega t)\cos(\Omega t')\langle 0|E_{y}\left(t,\mathbf{x}\right)E_{x}\left(t',\mathbf{x}'\right)|0\rangle\nonumber\\
&&+ \sin(\Omega t)\sin(\Omega t')\langle 0|E_{y}\left(t,\mathbf{x}\right)E_{y}\left(t',\mathbf{x}'\right) |0\rangle\;,
\end{eqnarray}
\begin{align}
\langle 0|\mathcal{E}_{\phi}\left(t,\mathbf{x}\right)\mathcal{E}_{\phi}\left(t',\mathbf{x}'\right)|0\rangle
=&\gamma^{-2}\left[\cos(\Omega t)\cos(\Omega t')\langle 0|E_{y}\left(t,\mathbf{x}\right)E_{y}\left(t',\mathbf{x}'\right) |0\rangle
-\cos(\Omega t)\sin(\Omega t')\langle 0|E_{y}\left(t,\mathbf{x}\right)E_{x}\left(t',\mathbf{x}'\right) |0\rangle\right.\nonumber\\
&\left.-\sin(\Omega t)\cos(\Omega t')\langle 0|E_{x}\left(t,\mathbf{x}\right)E_{y}\left(t',\mathbf{x}'\right) |0\rangle
+\sin(\Omega t)\sin(\Omega t')\langle 0|E_{x}\left(t,\mathbf{x}\right)E_{x}\left(t',\mathbf{x}'\right) |0\rangle\right]\;,
\end{align}
\begin{eqnarray}
\langle 0|\mathcal{E}_{z}\left(t,\mathbf{x}\right)\mathcal{E}_{z}\left(t',\mathbf{x}'\right)|0\rangle
&=& \langle 0|E_{z}\left(t,\mathbf{x}\right)E_{z}\left(t',\mathbf{x}'\right) |0\rangle
- \Omega R\sin(\Omega t')\langle 0|E_{z}\left(t,\mathbf{x}\right)B_{y}\left(t',\mathbf{x}'\right) |0\rangle\nonumber\\
&&- \Omega R\cos(\Omega t')\langle 0|E_{z}\left(t,\mathbf{x}\right)B_{x}\left(t',\mathbf{x}'\right) |0\rangle
- \Omega R\sin(\Omega t)\langle 0|B_{y}\left(t,\mathbf{x}\right)E_{z}\left(t',\mathbf{x}'\right) |0\rangle\nonumber\\
&&+\Omega^2 R^2 \sin(\Omega t)\sin(\Omega t')\langle 0|B_{y}\left(t,\mathbf{x}\right)B_{y}\left(t',\mathbf{x}'\right) |0\rangle\nonumber\\
&&+\Omega^2 R^2 \sin(\Omega t)\cos(\Omega t')\langle 0|B_{y}\left(t,\mathbf{x}\right)B_{x}\left(t',\mathbf{x}'\right) |0\rangle\nonumber\\
&&- \Omega R\cos(\Omega t)\langle 0|B_{x}\left(t,\mathbf{x}\right)E_{z}\left(t',\mathbf{x}'\right) |0\rangle
+\Omega^2 R^2 \cos(\Omega t)\sin(\Omega t')\langle 0|B_{x}\left(t,\mathbf{x}\right)B_{y}\left(t',\mathbf{x}'\right) |0\rangle\nonumber\\
&&+\Omega^2 R^2 \cos(\Omega t)\cos(\Omega t')\langle 0|B_{x}\left(t,\mathbf{x}\right)B_{x}\left(t',\mathbf{x}'\right) |0\rangle\;,
\end{eqnarray}
where
\begin{eqnarray}
&&\left\langle 0\left|E_{l}(t,\mathbf{x}) E_{p}\left(t',\mathbf{x}'\right)\right| 0\right\rangle=\frac{\hbar}{8 \pi \epsilon_{0} V}\int_{0}^{2 \pi} d\varphi \int_{0}^{\pi} \sin\theta d\theta \int_{0}^{\infty} d\omega_{k} \rho(\omega_k) \frac{\omega_k}{2} \left(\delta_{lp}-\frac{k_{l}k_{p}}{\boldsymbol{k}^{2}}\right) e^{-i( \omega_{k} t_{-}-\boldsymbol{k} \cdot \boldsymbol{R})}\label{correlationfunctionEE}\;,\\
&&\left\langle 0\left|B_{l}(t,\mathbf{x}) B_{p}\left(t',\mathbf{x}'\right)\right| 0\right\rangle=\frac{\hbar}{8 \pi\epsilon_{0} V}\int_{0}^{2 \pi} d\varphi \int_{0}^{\pi} \sin\theta d\theta \int_{0}^{\infty} d\omega_{k} \rho(\omega_k) \frac{\omega_k}{2c^2} \left(\delta_{lp}-\frac{k_{l}k_{p}}{\boldsymbol{k}^{2}}\right)  e^{-i( \omega_{k} t_{-}-\boldsymbol{k} \cdot\boldsymbol{R})}\label{correlationfunctionBB}\;,\\
&&\left\langle 0\left|E_{l}(t,\mathbf{x}) B_{p}\left(t',\mathbf{x}'\right)\right| 0\right\rangle=\frac{\hbar}{8 \pi \epsilon_{0} V}\int_{0}^{2 \pi} d\varphi \int_{0}^{\pi} \sin\theta d\theta\int_{0}^{\infty} d\omega_{k} \rho(\omega_k) \frac{\omega_k}{2c} \epsilon_{lpq} \frac{k_{q}}{\lvert \boldsymbol{k}\rvert} e^{-i( \omega_{k} t_{-}-\boldsymbol{k} \cdot\boldsymbol{R})}\label{correlationfunctionEB}\;,\\
&&\left\langle 0\left|B_{l}(t,\mathbf{x}) E_{p}\left(t',\mathbf{x}'\right)\right| 0\right\rangle=\frac{\hbar}{8 \pi \epsilon_{0} V}\int_{0}^{2 \pi} d\varphi \int_{0}^{\pi} \sin\theta d\theta\int_{0}^{\infty} d\omega_{k} \rho(\omega_k) \frac{\omega_k}{2c} \left(-\epsilon_{lpq} \frac{k_{q}}{\lvert \boldsymbol{k}\rvert}\right)  e^{-i( \omega_{k} t_{-}-\boldsymbol{k} \cdot\boldsymbol{R})}\label{correlationfunctionBE}\;.
\end{eqnarray}
Here $l,p,q=x,y,z$, $\epsilon_{lpq}$ denotes the Levi-Civita symbol, $\boldsymbol{R}=\mathbf{x}(t)-\mathbf{x}(t^{\prime})$ is the displacement vector along the atomic trajectory, and $t_-=t-t'$. The quantities $\omega_k$ and $\mathbf{k}$ denote the mode frequency and wave vector, respectively. Throughout the paper, the spectral density $\rho(\omega_k)$ is modeled by a Lorentzian distribution.
\end{widetext}


\begin{thebibliography}{00}
\bibitem{Fulling73}  S. A. Fulling, Nonuniqueness of canonical field quantization in Riemannian space-time, \href{https://link.aps.org/doi/10.1103/PhysRevD.7.2850}{Phys. Rev. D {\bf7}, 2850 (1973)}.
\bibitem{Davies75}  P. C. W. Davies, Scalar production in Schwarzschild and Rindler metrics, \href{https://dx.doi.org/10.1088/0305-4470/8/4/022}{J. Phys. A: Math. Gen. {\bf8}, 604 (1975)}.
\bibitem{Unruh76}  W. G. Unruh, Notes on black-hole evaporation, \href{https://link.aps.org/doi/10.1103/PhysRevD.14.870}{Phys. Rev. D {\bf14}, 870 (1976)}.
\bibitem{Hawking74}
S. Hawking,
Black hole explosions?,
\href{https://doi.org/10.1038/248030a0}{Nature {\bf 248}, 30 (1974)}.
\bibitem{Hawking75}  S. W. Hawking, Particle creation by black holes, \href{https://doi.org/10.1007/BF02345020}{Commun. Math. Phys. {\bf43}, 199 (1975)}.
\bibitem{Lamb47}  W. E. Lamb and R. C. Retherford, Fine structure of the Hydrogen atom by a microwave method, \href{https://link.aps.org/doi/10.1103/PhysRev.72.241}{Phys. Rev. {\bf72}, 241 (1947)}.
\bibitem{Bezginov19}  N. Bezginov, T. Valdez, M. Horbatsch, A. Marsman, A. C. Vutha, and E. A. Hessels, A measurement of the atomic hydrogen Lamb shift and the proton charge radius, \href{https://doi.org/10.1126/science.aau7807}{Science {\bf365}, 1007 (2019)}.
\bibitem{Audretsch95A}  J. Audretsch and R. M\"uller, Radiative energy shifts of an accelerated two-level system, \href{https://link.aps.org/doi/10.1103/PhysRevA.52.629}{Phys. Rev. A {\bf52}, 629 (1995)}.
\bibitem{Passante98}  R. Passante, Radiative level shifts of an accelerated hydrogen atom and the Unruh effect in quantum electrodynamics, \href{https://link.aps.org/doi/10.1103/PhysRevA.57.1590}{Phys. Rev. A {\bf57}, 1590 (1998)}.
\bibitem{Letaw1980}
J. R. Letaw and J. D. Pfautsch, Quantized scalar field in rotating coordinates, \href{https://doi.org/10.1103/PhysRevD.22.1345}{Phys. Rev. D {\bf 22}, 1345 (1980)}.
\bibitem{Bell83}
J. S. Bell and J. M. Leinaas, Electrons as accelerated thermometers, \href{https://doi.org/10.1016/0550-3213(83)90601-6}{Nucl. Phys. {\bf B212}, 131 (1983)}.
\bibitem{Hacyan1986}
S. Hacyan, A. Sarmiento, Vacuum energy of the electromagnetic field in a rotating system, \href{https://doi.org/10.1016/0370-2693(86)90582-4}{Phys. Lett. B {\bf 179}, 287 (1986)}.
\bibitem{Bell87}
J. S. Bell and J. M. Leinaas, The Unruh effect and quantum fluctuations of electrons in storage rings, \href{https://doi.org/10.1016/0550-3213(87)90047-2}{Nucl. Phys. {\bf B284}, 488 (1987)}.
\bibitem{Kim1987}
S. K. Kim, K. S. Soh, and J. H. Yee, Zero-point field in a circular-motion frame, \href{https://doi.org/10.1103/PhysRevD.35.557}{Phys. Rev. D {\bf 35}, 557 (1987)}.
\bibitem{Unruh98}
W. G. Unruh, Acceleration  for orbiting electrons, \href{https://doi.org/10.1016/S0370-1573(98)00068-4}{Phys. Rep. {\bf 307}, 163 (1998)}.
\bibitem{Audretsch95CQG}  J. Audretsch, R. M\"uller, and M. Holzmann, Generalized Unruh effect and Lamb shift for atoms on arbitrary stationary trajectories, \href{https://iopscience.iop.org/article/10.1088/0264-9381/12/12/010}{Class. Quantum Grav. {\bf12}, 2927 (1995)}.
\bibitem{yu261}
Y. Peng, J. Hu, and H. Yu,
Significant modifications of Lamb shift at small centripetal accelerations,
\href{https://doi.org/10.48550/arXiv.2603.05945}{arXiv:2603.05945}.
\bibitem{Purcell46}  E. M. Purcell, Spontaneous emission probabilities at radio frequencies, \href{https://doi.org/10.1007/978-1-4615-1963-8_40}{Phys. Rev. {\bf69}, 681 (1946)}.
\bibitem{Scully03}  M. O. Scully, V. V. Kocharovsky, A. Belyanin, E. Fry, and F. Capasso, Enhancing acceleration  from ground-state atoms via cavity quantum electrodynamics, \href{https://link.aps.org/doi/10.1103/PhysRevLett.91.243004}{Phys. Rev. Lett. {\bf91}, 243004 (2003)}.
\bibitem{Lochan20}  K. Lochan, H. Ulbricht, A. Vinante, and S. K. Goyal, Detecting acceleration-enhanced vacuum fluctuations with atoms inside a cavity, \href{https://doi.org/10.1103/PhysRevLett.125.241301}{Phys. Rev. Lett. {\bf 125}, 241301 (2020)}.
\bibitem{Stargen22}  D. Jaffino Stargen and K. Lochan, Cavity optimization for Unruh effect at small accelerations, \href{https://link.aps.org/doi/10.1103/PhysRevLett.129.111303}{Phys. Rev. Lett. {\bf129}, 111303 (2022)}.
\bibitem{Zheng25}
H.-T. Zheng, X.-F. Zhou, G.-C. Guo, and Z.-W. Zhou,
Enhancing analog Unruh effect via superradiance in a cylindrical cavity,
\href{https://link.aps.org/doi/10.1103/PhysRevResearch.7.013027}{Phys. Rev. Res. {\bf 7}, 013027 (2025)}.
\bibitem{yu26}
Y. Peng, Y. Zhou, J. Hu, and H. Yu,
Extensive Manipulation of Transition Rates and Substantial Population Inversion of Rotating Atoms Inside a Cavity,
\href{https://link.aps.org/doi/10.1103/6nqb-tnp6}{Phys. Rev. Lett. {\bf 136}, 013202 (2026)}.
\bibitem{Arya23}
N. Arya and S. K. Goyal,
Lamb shift as a witness for quantum noninertial effects,
\href{https://link.aps.org/doi/10.1103/PhysRevD.108.085011}{Phys. Rev. D {\bf 108}, 085011 (2023)}.
\bibitem{Arya24}
N. Arya, D. Jaffino Stargen, K. Lochan, and S. K. Goyal,
Strong noninertial radiative shifts in atomic spectra at low accelerations,
\href{https://link.aps.org/doi/10.1103/PhysRevD.110.085007}{Phys. Rev. D {\bf 110}, 085007 (2024)}.
\bibitem{sahota2604}
H. S. Sahota, S. Kaushal, and K. Lochan,
Cavity-controlled inhibition of decoherence in accelerated quantum detectors,
\href{https://doi.org/10.48550/arXiv.2604.02422}{arXiv:2604.02422}.
\bibitem{breuer07} H.-P. Breuer and F. Petruccione, \textit{The Theory of Open Quantum Systems} (Oxford University Press, Oxford, 2007).

\bibitem{footnote} The cavity mode considered here possesses a Lorentzian linewidth $\kappa=\omega_c/Q$, corresponding to a correlation time $\tau_B\sim\kappa^{-1}$. Throughout this work we assume the weak-coupling regime $g\ll\kappa$, where $g=d\sqrt{\omega_c/(2\hbar\epsilon_0 V)}$ is the atom--field coupling strength. Under this condition $\tau_B\sim\kappa^{-1}\ll g^{-1}$, so the cavity memory time is much shorter than the characteristic atomic evolution time and the Markov approximation is justified.
\end{thebibliography}
\end{document}